\documentclass[aps,prl,twocolumn,groupedaddress,superscriptaddress,showpacs,msmath,amssymb]{revtex4}

\usepackage{graphicx}
\usepackage{dcolumn}
\usepackage{bm}
\usepackage[usenames]{color}
\usepackage[normalem]{ulem}

\begin{document}

\title{
Observation of a Highly Spin Polarized Topological Surface State in GeBi$_{2}$Te$_{4}$
}

\author{K. Okamoto}
\affiliation{%
Graduate School of Science, Hiroshima University, 1-3-1 Kagamiyama, Higashi-Hiroshima 739-8526, Japan\\
}

\author{K. Kuroda}
\affiliation{%
Graduate School of Science, Hiroshima University, 1-3-1 Kagamiyama, Higashi-Hiroshima 739-8526, Japan\\
}

\author{H. Miyahara}
\affiliation{%
Graduate School of Science, Hiroshima University, 1-3-1 Kagamiyama, Higashi-Hiroshima 739-8526, Japan\\
}

\author{K. Miyamoto}
\affiliation{
Hiroshima Synchrotron Radiation Center, Hiroshima University, 2-313 Kagamiyama, Higashi-Hiroshima 739-0046, Japan\\
}

\author{T. Okuda}
\affiliation{
Hiroshima Synchrotron Radiation Center, Hiroshima University, 2-313 Kagamiyama, Higashi-Hiroshima 739-0046, Japan\\
}

\author{Z. S. Aliev}
\affiliation{%
Baku State University, General and Inorganic Chemistry Department, AZ1148 Baku, Azerbaijian\\
}

\author{M. B. Babanly}
\affiliation{%
Baku State University, General and Inorganic Chemistry Department, AZ1148 Baku, Azerbaijian\\
}

\author{I. R. Amiraslanov}
\affiliation{%
Institute of Physics, Azerbaijian National Academy of Science, AZ1143 Baku, Azerbaijian\\
}

\author{K. Shimada}
\affiliation{
Hiroshima Synchrotron Radiation Center, Hiroshima University, 2-313 Kagamiyama, Higashi-Hiroshima 739-0046, Japan\\
}

\author{H. Namatame}
\affiliation{
Hiroshima Synchrotron Radiation Center, Hiroshima University, 2-313 Kagamiyama, Higashi-Hiroshima 739-0046, Japan\\
}

\author{M. Taniguchi}
\affiliation{%
Graduate School of Science, Hiroshima University, 1-3-1 Kagamiyama, Higashi-Hiroshima 739-8526, Japan\\
}
\affiliation{
Hiroshima Synchrotron Radiation Center, Hiroshima University, 2-313 Kagamiyama, Higashi-Hiroshima 739-0046, Japan\\
}

\author{E. V. Chulkov}
\affiliation{%
Departamento de F\'{\i}sica de Materiales UPV/EHU
and Centro de F\'{\i}sica de Materiales CFM
and Centro Mixto CSIC-UPV/EHU,
20080 San Sebasti\'an/Donostia, Basque Country, Spain
\\
}
\affiliation{%
Donostia International Physics Center (DIPC),
             20018 San Sebasti\'an/Donostia, Basque Country,
             Spain\\
}

\author{A. Kimura}
\email{akiok@hiroshima-u.ac.jp}
\affiliation{%
Graduate School of Science, Hiroshima University, 1-3-1 Kagamiyama, Higashi-Hiroshima 739-8526, Japan\\
}

\date{\today}

\begin{abstract}
Spin polarization of a topological surface state for GeBi$_2$Te$_4$, the newly discovered three-dimensional topological insulator, has been studied by means of the state of the art spin- and angle-resolved  photoemission spectroscopy. It has been revealed that the disorder in the crystal has a minor effect on the surface state spin polarization and it exceeds $75\%$ near the Dirac point in the bulk energy gap region ($\sim$180~meV).
This new finding for GeBi$_{2}$Te$_{4}$ promises not only to realize a highly spin polarized surface isolated transport but to add new functionality to its thermoelectric and thermomagnetic properties.
\end{abstract}


\maketitle

Topological insulators (TIs) have recently emerged as a new state of quantum
matter, which are distinguished from conventional insulators by a massless Dirac cone surface state in the bulk energy gap, 
the so called topological surface state (TSS). 
The spin orientation of the TSS is locked with respect to
crystal momentum, resulting in a helical spin texture~\cite{FKM_07,FK_07}. 
The unique properties of topological surface electrons provide a fertile ground to realize new 
electronic phenomena, such as a magnetic monopole arising from the topological 
magneto-electric effect and Majorana fermions at the interface with a superconductor
~\cite{Hasan&Kane_RMP,Qi&Zhang_RMP}. Due to time-reversal symmetry, a TSS is 
protected from backscattering in the presence of a weak 
perturbation, a feature which is required for the realization of dissipationless spin 
transport in the absence of external magnetic fields in novel quantum devices~\cite{Xue_11,Xiu_11}.

A number of materials that hold spin-polarized TSSs
have been intensively studied, such as
Bi$_{1-x}$Sb$_{x}$~\cite{Hsieh_Science_09, Nishide_PRB_10},
Bi$_{2}$Se$_{3}$~\cite{Xia_NatPhys_09,
Kuroda_PRL_10_Bi2Se3,Bi2Se3_eph,Kim_PRL_11}, 
Bi$_{2}$Te$_{3}$~\cite{Chen_Science_09,Hsieh_PRL2009}
and thallium- and lead- based ternary compounds
\cite{eremeev1,Eremeev_Tl_PRB2011,Kuroda_PRL_10_TlBiSe2,Yan_EPL_10,Lin_PRL_10,Sato_PRL_10,Chen_PRL_10,Kuroda_PRL_12, Souma_PRL_12}.
Among these materials, Bi$_{2}$Se$_{3}$ has been regarded as the most promising 3D TI because it possesses a single TSS in a rather wide bulk energy gap~\cite{Xia_NatPhys_09, Kuroda_PRL_10_Bi2Se3}.
However, no surface isolated conduction has been observed for this binary compound~\cite{Analytis_PRB_10, Eto_PRB_10, Butch_PRB_10} .


\begin{figure} [t]
\includegraphics{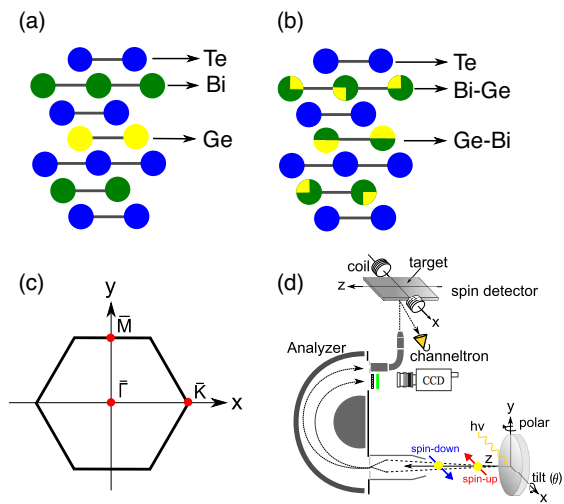}
\caption{(color online) 
(a) Ideal and (b) experimentally determined seven-layer blocks in of GeBi$_{2}$Te$_{4}$ crystal (see text).
(c) Surface Brillouin zone. 
(d) Experimental geometry for spin-ARPES measurement.
}
\end{figure}

\begin{figure} [t]
\includegraphics{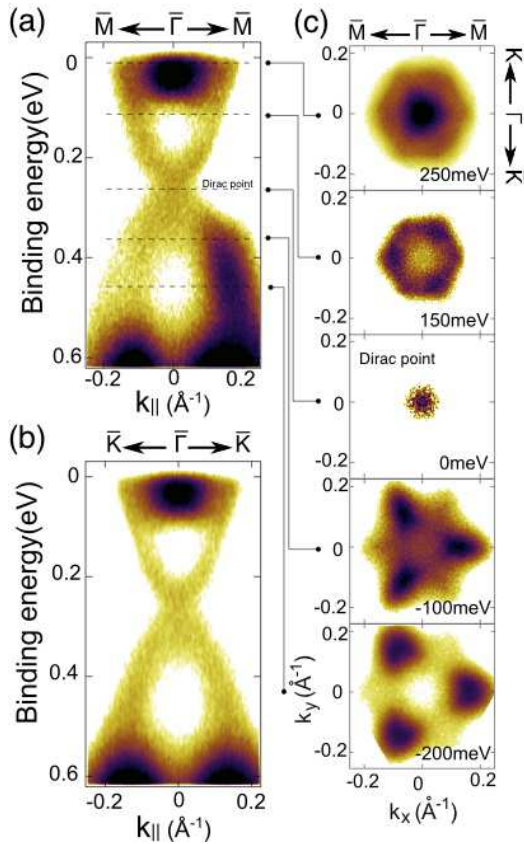}
\caption{(color online) (a) Experimental ARPES results for GeBi$_{2}$Te$_{4}$.
Energy dispersion curve along the (a) $\bar{\Gamma}\bar{M}$ and (b) $\bar{\Gamma}\bar{K}$ lines.
(c) Constant energy surfaces at 250~meV, 150~meV, 0~meV, -100~meV, -200~meV with respect to the Dirac point ($E_{\rm B}$=260~meV).
}
\end{figure}

\begin{figure*} [t]
\includegraphics{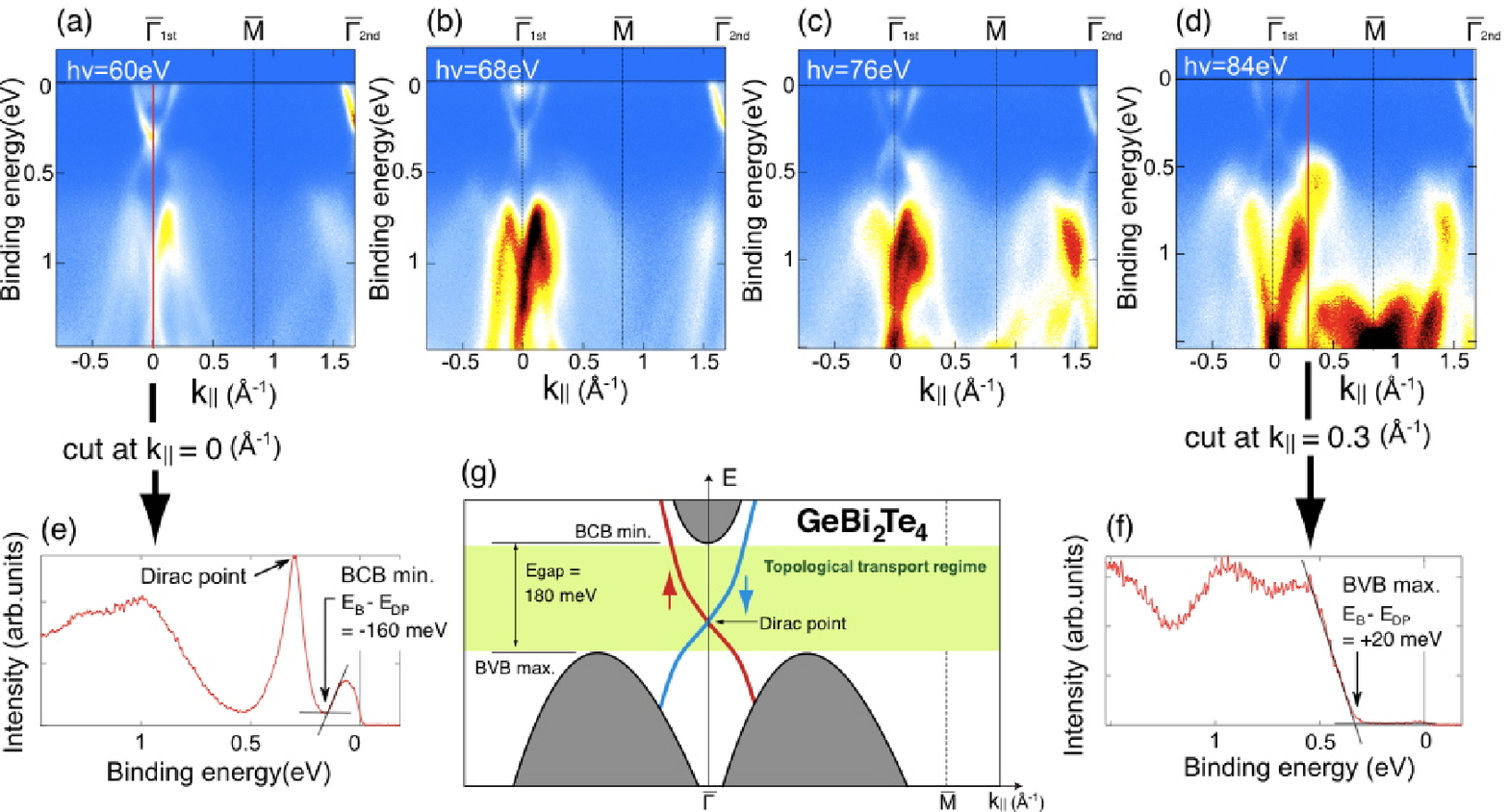}
\caption{(color online) ARPES $E$-$k_{\parallel}$ map over a wide $k_{\parallel}$ range along the $\bar{M}$$\bar{\Gamma}$$\bar{M}$ line acquired at $h\nu$= (a) 60, (b) 68, (c) 76 and (d) 84~eV.
Energy distribution curves in the $E_{\rm B}$ range of 0-1.4~eV sliced along the constant $k_{\parallel}$ lines at (e) 0~\AA$^{-1}$ for $h\nu$=60~eV and (f) 0.3~\AA$^{-1}$ for $h\nu$=84~eV.
(g) Schematics of surface (blue and red lines) and bulk band structures (gray shaded area) of GeBi$_{2}$Te$_{4}$ figured out from the present experimental results.}
\end{figure*}

A homologous series of pseudobinary compounds $n$GeTe-$m$Bi$_{2}$Te$_{3}$ was intensively studied in terms of their thermoelectric, galvano- and thermomagnetic properties~\cite{Shelimova_00, Shelimova_01, Matsunaga_10}.
Among them, GeBi$_{2}$Te$_{4}$ was theoretically proposed as a member of the 3D TIs~\cite{Menshchikova_JETP_11, Eremeev_NC_12, Yang_NM_12}.
It was experimentally verified to be a 3D TI possessing a single TSS by an angle-resolved photoemission spectroscopy (ARPES) experiment~\cite{Neupane_12}.
The crystal structure of GeBi$_{2}$Te$_{4}$ assumed in the calculation was composed of seven-layer (7L) blocks formed by the atomic layer sequence Te-Bi-Te-Ge-Te-Bi-Te as shown in Fig.1(a).
However, the structure in the {\it real} material was found to deviate from the ideal one.
It was revealed by an X-ray diffraction study that the central cation layer of the 7L block is not pure Ge, but contains equal amounts of randomly distributed Ge and Bi atoms and the other two cation layers result also in a substantial intermixing~\cite{Karpinsky_98} [Fig.1(b)].
This observation prompts the important question of how the spin polarization of the TSS 
would be affected by the intermixing in the GeBi$_{2}$Te$_{4}$ crystal.

Revealing the size of the bulk energy gap and the $\mathbf k$-space location of the TSS especially with respect to the bulk band gap is crucial for realizing topological transport with a sufficiently isolated surface conduction.
Also, the surface spin polarization needs to be as high as possible even though its magnitude is predicted to be reduced to 50\%-60\%~for Bi$_{2}$Se$_{3}$ and Bi$_{2}$Te$_{3}$ due to inevitable spin and orbital entanglement~\cite{Yazyev_10}.
Although the topological surface state was experimentally identified~\cite{Neupane_12}, such important information on these aspects in the presence of mixed interlayers is so far missing for GeBi$_{2}$Te$_{4}$.
Here, we report that the Dirac point of the TSS is located within the bulk band gap of $\sim180$~meV, and the TSS has a substantial spin polarization above 75\%, which are revealed by means of spin-resolved/integrated ARPES.
This finding promises to realize a surface isolated highly spin polarized transport and add new functionality to its thermoelectric and thermomagnetic properties.

A single crystalline ingot of GeBi$_{2}$Te$_{4}$ was grown by the vertical Bridgman-Stockbarger method.
The grown crystal was characterized by X-ray diffraction using a Bruker D8 ADVANCE diffractometer with Cu K$_{\alpha}$ radiation.
Spin-integrated ARPES measurement was carried out with synchrotron radiation at the linear undulator beamline (BL-1) of Hiroshima Synchrotron Radiation Center (HiSOR).
The spin-resolved ARPES (SARPES) experiment was performed with a He discharge lamp at the Efficient SPin Resolved SpectroScOpy  (ESPRESSO) end station with the VLEED-type spin polarimeter~\cite{Okuda_RSI_11}.
The spin polarimeter utilizes [Fig.1(d)] a magnetic target of a Fe(001)-p(1$\times$1)-O film grown on MgO(001) substrate, which achieves a 100 times higher efficiency compared to those of conventional Mott-type spin detectors~\cite{Okuda_RSI_11}.
Photoelectron spin polarizations were measured by switching the direction of in-plane target magnetizations.
This simultaneously eliminate any instrumental asymmetry, which is a great advantage for the quantitative spin analysis of non-magnetic systems, as in the present case.
The angle of light incidence was $50^{\circ}$ relative to the lens axis of the electron analyzer.
The sign of the polar (tilt)  angle is defined as positive, in the case of a clockwise (anticlockwise) rotation about y-axis (x-axis) as shown in Fig.1(d).
The energy- and wavenumber- resolutions for the synchrotron radiation ARPES (BL-1) were set to better than 48~meV and 0.05~$\rm \AA^{-1}$, respectively, while those for the ARPES (SARPES) with He discharge lamp were set to 19~meV and $<$ 0.036$\rm \AA^{-1}$ (27~meV and $<$ 0.06 $\rm \AA^{-1}$).
The measurement temperatures at BL-1 and at ESPRESSO end station were 10 and 50~K, respectively.  
The samples were cleaved {\it in-situ} under an ultrahigh vacuum below $1\times 10^{-8}$ Pa.

\begin{figure*} [t]
\includegraphics{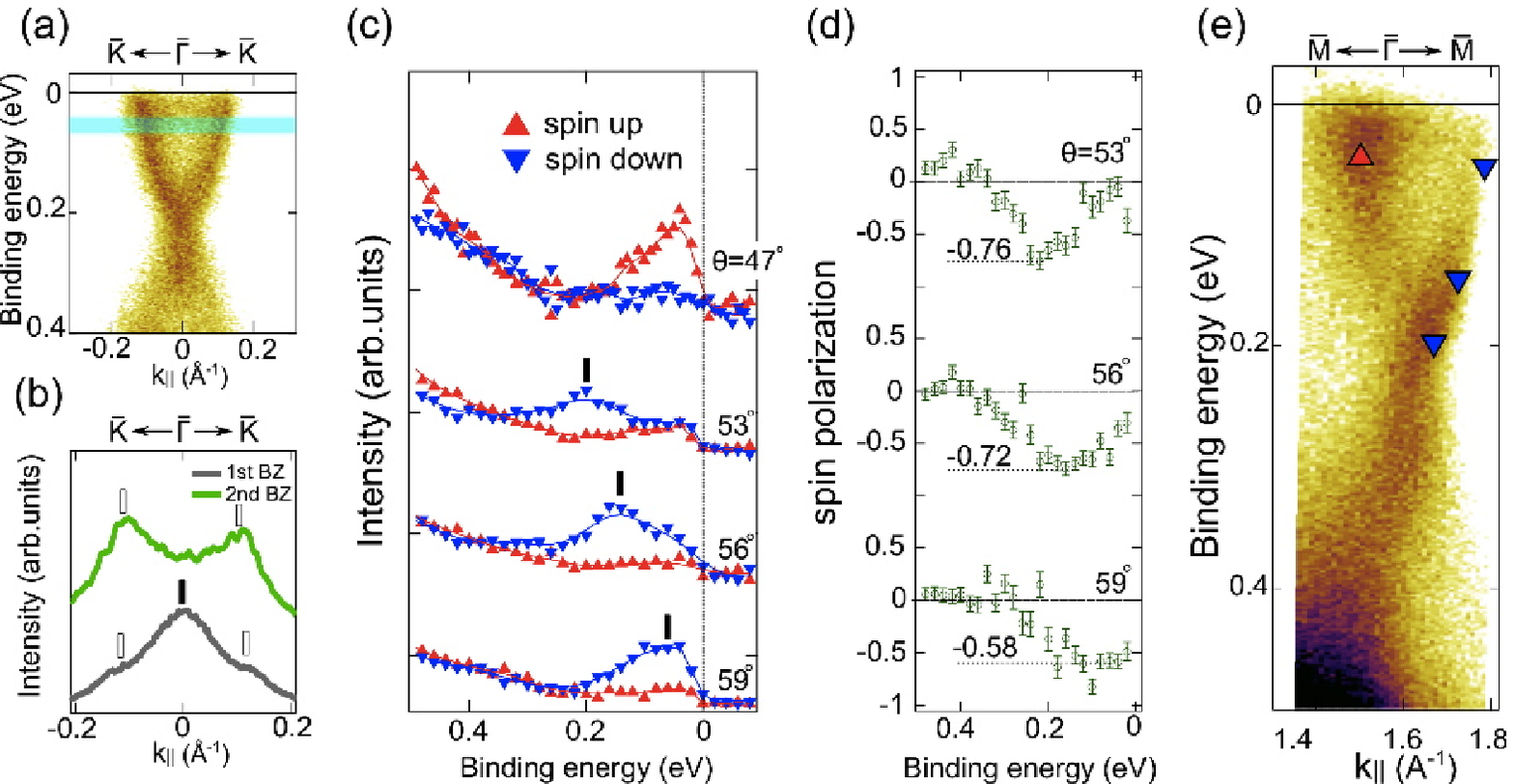}
\caption{(color online) (a) Spin integrated energy dispersion curve along $\bar{K}\bar{\Gamma}\bar{K}$ in the 2nd SBZ. 
(b) Momentum distribution curves at $E_{\rm B}$=70~meV for 1st ({\it bottom}) and 2nd ({\it top}) SBZs. 
(c) Spin resolved energy distribution curves of GeBi$_{2}$Te$_{4}$ for emission angles ($\theta$'s) of 47$^{\circ}$, 53$^{\circ}$, 56$^{\circ}$ and 59$^{\circ}$ and corresponding spin polarizations for $\theta$=53$^{\circ}$, 56$^{\circ}$ and 59$^{\circ}$ are shown in panel (d).
(e) ARPES results in the 2nd SBZ. The contour plot has superimposed triangles pointing up and down indicating the spin character of the corresponding spectral features, as derived from spin-resolved spectra in panel (c).
}
\end{figure*}

Figures 2(a) and 2(b) show the ARPES energy dispersion curves along the $\bar{\Gamma}$$\bar{M}$ and $\bar{\Gamma}$$\bar{K}$ lines of the surface Brillouin zone (SBZ) [Fig.1(c)], respectively. 
Two surface energy bands, i.e. a topological surface state (TSS) with crossing point at a binding energy ($E_{\rm B}$) of 260 meV (Dirac point), are clearly seen along these lines.
The bulk conduction band (BCB) is enclosed by the TSS and crosses the Fermi energy ($E_{\rm F}$) with a substantial photoemission intensity.
Constant energy contours in the ${\mathbf k}_\parallel$  range $-0.25$~\AA$^{-1}\le k_{x}, k_{y}\le +0.25$~\AA$^{-1}$
from -200 to +250~meV with respect to the Dirac point ($E_{\rm B}$=260~meV) are shown in Fig.2(c).
A hexagonally shaped constant energy contour is observed at $E_{\rm F}$, whose shape is preserved even at $E_{\rm B}$=150~meV.
The hexagon of the TSS evolves into the point-like feature at the Dirac point and is again strongly deformed into a snow-flake below the Dirac point.
Another triangular feature is enclosed within the TSS at $E_{\rm F}$, which comes from the bulk conduction band.
These features are consistent with the previous ARPES experiment~\cite{Neupane_12}.
Here it has to be mentioned that the size of constant energy surface of GeBi$_{2}$Te$_{4}$ ($|k_{x}|$$\sim$0.1~\AA$^{-1}$ at 150~meV above the Dirac point) is almost twice as large as that of the {\it ordered} Bi$_{2}$Se$_{3}$ ($\sim$0.05~\AA$^{-1}$ at the same energy)~\cite{Kuroda_PRL_10_Bi2Se3}.
This result implies that the intermixing of the GeBi$_{2}$Te$_{4}$ crystal would broaden the momentum width of the TSS. 

To determine the $\mathbf k$-space location of the bulk states with respect to the TSS, we have performed a detailed photon energy dependence study over a wide $k_{\parallel}$ range.
The ARPES measurements were performed with several incident photon energies ($h\nu$'s) from 60 to 84~eV to cover the whole Brillouin zone along the $k_{z}$ direction.
Figures 3(a)-(d) show the $E$-$k_{\parallel}$ map over a wide $k_{\parallel}$ range along the $\bar{M}$$\bar{\Gamma}$$\bar{M}$ line acquired at $h\nu$=60, 68, 76 and 84~eV.
The surface states at the $\bar{\Gamma}$ points in the 1st ($\bar{\Gamma}_{1st}$) and 2nd ($\bar{\Gamma}_{2nd}$) SBZs are found to be identical,
which signifies a single TSS in this compound.
The Dirac point energy does not change with $h\nu$ except for a time-dependent energy shift as will be discussed later, while the bulk states do, which again confirms their respective two- and three- dimensional nature.
At $h\nu$=60~eV, the BCB enclosed by the TSS is clearly identified.
In going to higher $h\nu$ it gradually shifts towards $E_{\rm F}$ and almost vanishes finally at $h\nu$=84~eV.
The bulk valence band (BVB), on the other hand, gradually grows up and shifts to lower $E_{\rm B}$ with increasing $h\nu$, achieving its maximum (minimum in $E_{\rm B}$) at $h\nu$=84~eV.

In Figs. 3(e) and 3(f) are shown the energy distribution curves in the $E_{\rm B}$ range of 0-1.4~eV sliced along the constant $k_{\parallel}$ lines at 0~\AA$^{-1}$ for $h\nu$=60~eV and 0.3~\AA$^{-1}$ for $h\nu$=84~eV.
In Fig.3(e), a sharp peak is observed at the Dirac point energy ($E_{\rm B}$=300~meV) and the BCB exhibits a Fermi energy cutoff accompanying a higher $E_{\rm B}$ tail.
Here, the BCB minimum is found at $E_{\rm B}$=140~meV (160~meV above the Dirac point) by extrapolating the higher energy tail to 'zero' intensity with a linear function.
To determine the BVB maximum, another EDC is given in Fig.3(f) and shows a monotonic decrease in intensity with decreasing $E_{\rm B}$.
By applying a similar fitting procedure to that used for the BCB, the BVB maximum energy is estimated to be $E_{\rm B}$=340~meV.
Since, as is commonly observed for Bi$_{2}$Se$_{3}$~\cite{King_PRL_11}, the 20~meV time-dependent energy shift to higher $E_{\rm B}$ occurs at the same time as that of the TSS, one may assume that the BVB maximum is located at 20~meV below the Dirac point. 
Thus these results lead to the conclusion that the total energy gap between the BVB maximum and BCB minimum is 180~meV in GeBi$_{2}$Te$_{4}$.
An important finding is that the Dirac point of TSS is located inside this indirect bulk energy gap (20~meV above the BVB maximum and 160~meV below the BCB minimum) as schematically shown in Fig.3(g).  

To unveil the spin characteristics of the TSS, the SARPES experiment has been carried out.
Two momentum distribution curves acquired at $E_{\rm B}$=70~meV near the $\bar{\Gamma}$ point in the 1st and the 2nd SBZs are compared in Fig.4(b).
In the figure, a significant overlap of the bulk conduction band intensity is recognized at the 1st SBZ, while the bulk-derived spectral intensity is well suppressed at the 2nd SBZ as can also be seen Fig.4(a). 
It is apparent that it would be better to choose the 2nd SBZ with larger emission angles for a quantitative spin analysis since the overlap of the TSS with the BCB can be avoided.
Figure 4(c) shows the spin-resolved energy distribution curves (EDCs) of GeBi$_{2}$Te$_{4}$ at emission angles of 47$^{\circ}$, 53$^{\circ}$, 56$^{\circ}$ and 59$^{\circ}$, where the respective $k_{\parallel}$ values are -0.13, 0.03, 0.09 and 0.15~\AA$^{-1}$ with respect to the $\bar{\Gamma}_{2nd}$ point ($k_{\parallel}$=1.66~\AA$^{-1}$).
Here, the spin-up and spin-down spectra are plotted with triangles pointing-up and -down, respectively.
At $\theta$=59$^{\circ}$, the spin-down intensity is predominant and crosses $E_{\rm F}$, while the spin-up intensity is quite small and featureless.
On the other hand, at $\theta$=47$^{\circ}$, which corresponds to another TSS branch, the spin-up intensity dominates with a quite small spin-down intensity near $E_{\rm F}$.
The observed anti-symmetric spin polarization at the two surface state branches are indeed a manifestation of a 3D TI.
The spin-down peak moves to higher $E_{\rm B}$ with decreasing $\theta$, which parallels the TSS dispersion in the bulk energy gap region [Fig.4(e)].
The spin polarizations at 59$^{\circ}$ is about 60\%~near $E_{\rm F}$ and it exceeds 70\% below 56$^{\circ}$ as shown in Fig.4(d).
Importantly, the spin polarization is further enhanced closer to the Dirac point and reaches 76\% at 53$^{\circ}$.
Since previous works on the surface state spin polarizations of the other 3D TIs were deduced only for the TSS outside the bulk energy gap region with rather large electron momenta, which might be due to insufficient instrumental angular resolutions, a direct comparison with the present result involving the value in the vicinity of the Dirac point might be difficult.
Nevertheless, we would say that the observed spin polarization of the TSS for GeBi$_{2}$Te$_{4}$ seems to be comparable that of the {\it ordered} Bi$_{2}$Se$_{3}$ ($\sim$75\%)~\cite{Pan_PRL_11}.
This indicates that the intermixing of the crystal has only a weak effect on the spin polarization of the TSS.

In conclusion, the size of the bulk energy gap for GeBi$_{2}$Te$_{4}$ is determined to be $\sim$180~meV and a topological surface state below and above the Dirac point is found to be isolated from the bulk band.
Importantly, it is revealed that the disorder in the GeBi$_{2}$Te$_{4}$ crystal has a minor effect on the magnitude of the surface state spin polarization and it indeed exceeds 75\%~in the bulk energy gap region.
This new finding promises to add novel functionality to the already known interesting thermoelectric and thermomagnetic properties of GeBi$_{2}$Te$_{4}$.

We thank J.~Jiang, H.~Hayashi, Y.~Nagata, and T.~Horike for their technical support in the ARPES measurement at BL-1 of Hiroshima Synchrotron Radiation Center (HiSOR).
This work was financially supported by Grant-in-Aid for Scientific Research Kiban A (No.23244066) and Kiban B (No.23340105) of the Japan Society for the Promotion of Science (JSPS).
The ARPES measurements were performed with the approval of the Proposal Assessing Committee of HSRC (Proposal No.12-A-24).

\end{document}